\documentstyle[epsfig]{mn}

\begin{document}

\title[] {Energisation of interstellar media and cosmic ray production by jets from X-ray binaries}
\author[Fender, Maccarone \& van Kesteren]  
{ R. P. Fender$^{1,2}$, T.J. Maccarone$^2$, Z. van Kesteren$^2$\\
$^1$School of Physics and Astronomy, University of Southampton
Hampshire SO17 1BJ \\
$^2$Astronomical Institute `Anton Pannekoek', University of
Amsterdam, Kruislaan 403, 1098 SJ Amsterdam, The Netherlands\\
}

\maketitle

\begin{abstract}
Drawing on recent estimates of the power of jets from X-ray binary
systems as a function of X-ray luminosity, combined with improved
estimates of the relevant Log(N)-Log(L$_{\rm X}$) luminosity
functions, we calculate the total energy input to the interstellar
medium (ISM) from these objects. The input of kinetic energy to the
ISM via jets is dominated by those of the black hole systems, in
contrast to the radiative input, which is dominated by accreting
neutron stars.  Summing the energy input from black hole jets $L_{\rm
J}$ in the Milky Way, we find that it is likely to correspond to $\geq
1$\% of $L_{\rm SNe}$, the time-averaged kinetic luminosity of
supernovae, and $\geq 5$\% of $L_{CR}$, the cosmic ray luminosity.
Given uncertainties in jet power estimates, significantly larger
contributions are possible.  Furthermore, in elliptical galaxies with
comparable distributions of low mass X-ray binaries, but far fewer
supernovae, the ratio $L_{\rm J} / L_{\rm SNe}$ is likely to be larger
by a factor of $\sim 5$.  We conclude that jets from X-ray binaries
may be an important, distributed, source of kinetic energy to the ISM
in the form of relativistic shocks, and as a result are likely to be a
major source of cosmic rays.
\end{abstract}



\section{Introduction}

The interstellar medium (ISM) in the Milky Way and other galaxies
receives a continuous input of energy in the form of both bulk motions
and radiation. The result of this energy input is the heating of the
ISM, the generation of interstellar turbulence and the production of
high energy cosmic rays. The most violent processes are those
associated with stellar evolution: stellar winds, protostellar jets
and supernovae (SNe). More specifically, cosmic ray production is only
likely to be associated with high velocity (i.e. $\ga 0.1c$) shocks,
and so their most likely source has long been considered to be the
SNe. See e.g. Chevalier (1977), McCray \& Snow (1979) and Elmegreen \&
Scalo (2004) for reviews of these processes.

However, in recent years it has become clear that there is at least
one other potential source of powerful relativistic shocks in the ISM,
namely the jets resulting from accretion onto stellar mass black holes
and neutron stars in X-ray binary systems (Mirabel \& Rodriguez 1999;
Fender 2005).  These jets seem to be powerful and ubiquitous in the
majority of such systems which exist in 'hard' X-ray spectral states
and/or are undergoing X-ray outbursts (Fender 2001; Fender, Belloni \&
Gallo 2004). Heinz \& Sunyaev (2002) estimated the total power input
from a handful of the most powerful jets in our galaxy and concluded
that they might contribute 1--10\% of the cosmic ray luminosity of the
galaxy, perhaps with a specific spectral signature. However, recent
advances in our estimates of the scaling of jet power with X-ray
luminosity (Fender, Gallo \& Jonker 2003; Fender, Belloni \& Gallo
2004), combined with improvements in our knowledge of the distribution
of X-ray sources as a function of luminosity (e.g. Grimm, Gilfanov \&
Sunyaev 2002,2003; Gilfanov 2004) have allowed us to recalculate more
accurately the sum of the power input from such jets. In this paper we
present the results of these calculations.

\begin{figure}
\centerline{\epsfig{file=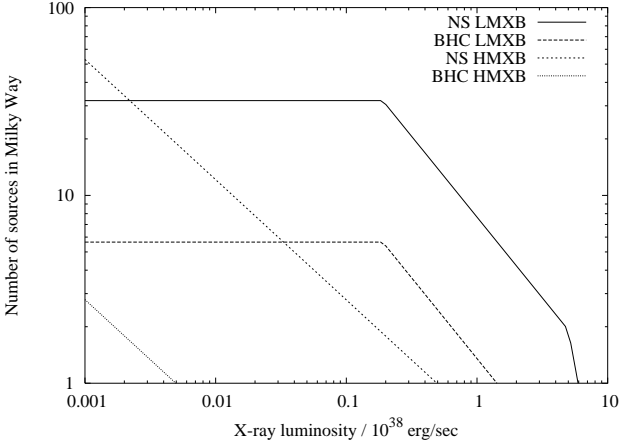, width=6cm, angle=270}}
\caption{Estimated total numbers of X-ray binaries in the Milky Way as
a function of X-ray luminosity. At the highest luminosities, above
$10^{35}$ erg s$^{-1}$, the low-mass X-ray binaries (LMXBs) dominate
the source population; although at even lower luminosities numbers of
high-mass X-ray binaries (HMXBs) may come to dominate, the numbers are
not well measured at those levels. Since the process of disc-jet
coupling does not seem to be strongly affected by the nature of the
companion star (see e.g. discussion in Fender 2005) this figure
illustrates that the total X-ray binary jet power into the
interstellar medium will be dominated by the LMXBs.}
\end{figure}

\section{Luminosity functions}

\subsection{Differential luminosity function for LMXBs}

Grimm et al. (2003) presented the X-ray luminosity function (XLF) for
LMXBs in the Milky Way in the form of a cut-off power law. Since then,
Gilfanov (2004) has produced an average differential XLF for LMXBs in
{\em all} galaxies. This function is in the form of a power law with
two breaks:

\begin{eqnarray}
\frac{dN}{dL_{X,38}}=\left\{ \begin{array}{ll}
\renewcommand{\arraystretch}{3}
K_1 \left(L_{X,38}/L_{b,1}\right) ^{-\alpha_1}
                        & \mbox{\hspace{0.9cm} $L_{X,38}<L_{b,1}$}\\
K_2 \left(L_{X,38}/L_{b,2}\right)^{-\alpha_2}
                        & \mbox{$L_{b,1}<L_{X,38}<L_{b,2}$}\\
K_3 \left(L_{X,38}/L_{cut}\right)^{-\alpha_3}
                        & \mbox{$L_{b,2}<L_{X,38}<L_{cut}$}\\
0                       & \mbox{\hspace{0.9cm} $L_{X,38}>L_{cut}$}\\
\end{array}
\right.
\label{eq:uxlf}
\end{eqnarray}
where $L_{X,38}=L_X/10^{38}$ erg s$^{-1}$ and normalizations $K_{1,2,3}$ are
related by
\begin{eqnarray}
K_2=K_1 \left(L_{b,1}/L_{b,2}\right)^{\alpha_2}\nonumber \\
K_3=K_2 \left(L_{b,2}/L_{cut}\right)^{\alpha_3}\nonumber
\end{eqnarray}

The best fit to the overall normalisation is given as

\begin{equation}
K_1 = 440.4 \pm 25.9 {\phantom{00}}{\rm per {\phantom{0}}10^{11} M_{\odot}}
\end{equation}

In the following we shall use this as the template LMXB XLF for all
galaxies. Based on Liu et al. (2000, 2001) we estimate the black hole
and neutron star fractions amongst the LMXBs to be $f_{BH} = 0.2$, $f_{NS}
= 0.8$. This luminosity function, normalised to the Milky Way using a
total stellar mass of $4.5 \times 10^{10}$ M$_{\odot}$ (Gilfanov 2004)
is plotted in Fig 1.

\subsection{Differential luminosity function for HMXBs}

For HMXBs, we use the differential form of the XLF from Grimm et
al. (2003), which is simply a steep power law:

\begin{equation}
\frac{dN}{dL_{X,38}} = K L_{\rm X,38}^{-1.64},
\end{equation}

where for the Milky Way $K \sim 0.7$. Based on Liu et al. (2000, 2001)
we estimate the black hole and neutron star fractions amongst the
HMXBs to be $f_{BH} = 0.05$, $f_{NS} = 0.95$. This function is plotted in
Fig 1 alongside the LMXB XLF.

\subsection{$L_{\rm J}(L_{\rm X})$ for X-ray binary jets}

\subsubsection{Steady ('low/hard' state) jets}

Corbel et al. (2000; 2003) and Gallo, Fender \& Pooley (2003) have found an
apparently universal correlation between radio and X-ray luminosities
for black hole candidate (BHC) binaries in the `low/hard' and
`quiescent' X-ray states (which seem to correspond to accretion rates
below about 1--10\% of the Eddington limit). The relation has the form

\begin{equation}
L_{\rm radio} \propto L_{\rm X}^{b}
\end{equation}

where $b \sim 0.7$. The relation between jet and radio luminosities
may also be considered to have a power-law form

\begin{equation}
L_{\rm radio} \propto L_{\rm J}^c
\end{equation}

Several models of steady, conical jets predict $c \sim 1.4$
(e.g. Blandford \& K\"onigl 1979; Heinz \& Sunyaev 2003), which leads to

\begin{equation}
L_{\rm J} = A_{\rm steady,BHC} L_{\rm X}^{0.5}
\end{equation}

in Eddington units, or

\begin{equation}
L_{\rm J,38} = A_{\rm steady,BHC} (L_{X,38})^{0.5} \left(\frac{M}{M_{\odot}}\right)^{0.5}
\end{equation}

indicating that the fractional jet power increases rapidly as the X-ray
luminosity decreases (Fender, Gallo \& Jonker 2003; Fender, Belloni \& Gallo 2004).

The value of the normalisation $A_{\rm steady,BHC}$ is uncertain. Fender, Gallo \&
Jonker (2003) estimated (in their opinion conservatively) that $A_{\rm
steady, BHC} \geq 6 \times 10^{-3}$. While it is certainly far from the
consensus, larger values of $A_{\rm steady,BHC}$ may be more realistic. Malzac, Merloni
\& Fabian (2004) have suggested that even at $L_{\rm X} \sim 10^{-3}
L_{\rm Edd}$, the jet may be an order of magnitude more powerful than
the X-ray emission, which corresponds to $A_{\rm steady,BHC} \sim
0.3$. Alternatively, we can set the value of $A_{\rm steady,BHC}$ so that it
corresponds to the transition from the low/hard to high/soft X-ray
states (see e.g. McClintock \& Remillard 2005 for a discussion), in
which case $A_{\rm BHC} \sim 0.1$. In all the following discussions we
shall consider $0.006 \leq A_{\rm BHC} \leq 0.3$ as covering the full
range of likely values.

We further assume that $A_{\rm steady, NS} = 0.1 A_{\rm steady, BHC}$,
based on the lower `radio loudness' of NS XRBs (Fender \& Kuulkers
2001; Migliari et al. 2003; Fender, Gallo \& Jonker 2003; Muno et
al. 2004; Migliari \& Fender in prep).  It should be stressed that the
$L_{\rm radio} \propto L_{\rm X}^{0.7}$ relation has {\em not} been
established for neutron star X-ray binaries, and is only assumed in
Fender, Gallo \& Jonker (2003) by analogy with the BHCs. However,
since the BHCs are so much more radio loud then this is not
significant for the total energy budget calculated here.

\subsubsection{$L_{\rm J}(L_{\rm X})$ for transient jets}

A fit to the power in transient optically thin ejection events from
{\em black hole} X-ray binaries as a function of X-ray luminosity
(Fender, Belloni \& Gallo 2004), simplified to one significant figure,
gives

\begin{equation}
L_{\rm J} = A_{\rm trans, BHC} L_{\rm X}^{0.5}
\end{equation}

in Eddington units, where $A_{\rm trans, BHC} \sim 0.4$. The same
$L_{\rm X}^{0.5}$ dependence (see Fender, Belloni \& Gallo 2004 for
caveats) as for the steady jets means the equation has the same form
with simply a different normalisation:

\begin{equation}
L_{\rm J,38} = A_{\rm trans,BHC} (L_{X,38})^{0.5} \left(\frac{M}{M_{\odot}}\right)^{0.5}
\end{equation}

It should be stressed that the normalisation is also rather uncertain,
being based on a small sample of sources for which the energy estimate
is itself intrinsically uncertain and different to that used for the
steady jet sources (see discussion in Fender, Belloni \& Gallo
2004). Nevertheless, we consider it likely that this estimate is
accurate to within one order of magnitude, and that if it is not then
it is more likely to be an underestimate than an overestimate. Based
upon this we will consider in this paper the range of likely values to
be $0.04 \leq A_{\rm trans,BHC} \leq 4.0$.

For the transient sources we further need to consider periods in
'soft' X-ray states when the radio emission, and hence presumably the
jet, are suppressed or 'quenched' (Fender et al. 1999; Corbel et al.
2001, 2004; Fender, Belloni \& Gallo 2004; Fender 2005).  The physical
processes responsible for the suppression of the jet are not clear,
but they seem to reduce the radio emission by a factor $\geq 30$ -- as
a result we shall consider that when in such states the jet power
drops to zero. The duty cycle in such states varies from outburst to
outburst and it is hard to estimate a reasonable mean value. Based
upon outbursts of the black hole LMXBs GX 339-4, XTE J1550-564 and XTE
J1859+226 it seems a value of 60-80\% for the duty cycle in 'soft'
jet-suppressed states is reasonable (J. Homan, private communication;
see also Fender, Belloni \& Gallo 2004; Homan \& Belloni 2005). As a
result, we multiply the integrated power from transients by a factor
0.3. While this number is rather uncertain, we note that it is still the
uncertainty in the value of the normalisation $A_{\rm trans}$ which
dominates the integrated power input.

In the absence of sufficient good data on neutron star transient
outburst with which to make a fit to $A_{\rm trans, NS}$, we simply follow
the approach adopted for the steady jets, i.e. $A_{\rm trans, NS} = 0.1
A_{\rm trans, BHC}$.

\section{Total power}

As noted above, $P_{\rm J}(L_{\rm X})$ is probably a function both of
X-ray state and of the nature of the accreting compact star (neutron
star or black hole). Convolved with both HMXB and LMXB luminosity
functions, this presents us with eight populations of jet producing
objects to consider (LMXB/HMXB $\times$ steady/transient $\times$
NS/BHC).

The integrated total jet power for any one of these classes is given by

\begin{equation}
L_{\rm J, total} = f \int^{L_{\rm X, max}}_{L_{\rm X, min}} P_{\rm J} \frac{dN}{dL_{\rm X}} dL_{\rm X}
\end{equation}

where $P_{\rm J}(L_{\rm X})$ is the jet power at a given X-ray
luminosity $L_{\rm X}$, and $N_{\rm X} dL_{\rm X}$ is the differential
luminosity function for the class of object. The term $f$ is a
correction factor which can be separated into two parts:

\begin{equation}
f = f_{pop} f_{spec}
\end{equation}

The first term, $f_{pop}$ is simply the fraction of the population of
that class of objects, in the sense that $f_{pop, NS} + f_{pop, BHC} =
1$. The second term, $f_{spec}$ compensates for the fact that the
X-ray luminosity functions given above, from Grimm et al. (2003) and
Gilfanov (2004) consider the X-ray luminosity in the range $\leq 10$
keV, whereas the expressions for the jet power were arrived at
considering the total X-ray power. This term is therefore in some
sense a bolometric correction. The values of $f_{spec}$ for black
holes are estimated in Portegies Zwart, Dewi \& Maccarone (2004) for
the black hole candidates.  For the neutron star sources we followed
essentially the same method. Estimated values for $f_{pop}$ and
$f_{spec}$ are presented in table 1.

\begin{table*}
\caption{Estimated values for the population $f_pop$ and spectral $f_spec$
corrections to be made to the integrals for the different classes of X-ray binary, 
as well as the jet power normalisation $A$. See text for more details.} 
\begin{tabular}{cccccccc}
& & \multicolumn{3}{c}{steady jets} & \multicolumn{3}{c}{transient jets} \\
& & $f_{pop}$ & $f_{spec}$ & A & $f_{pop}$ & $f_{spec}$ & A \\
LMXB & NS & 0.85 & 5.0 & 0.0006--0.03 & 0.85 & 2.0 & 0.0004--0.4\\
     & BHC & 0.15 & 10.5 & 0.006--0.3 & 0.15 & 1.2 & 0.04--4.0 \\
HMXB & NS & 0.95 & 5.0 & 0.0006--0.03 & 0.95 & 2.0 & 0.0 \\
     & BHC & 0.05 & 10.0 & 0.006--0.3 & 0.05 & 1.2 & 0.04--4.0 \\
\end{tabular}
\end{table*}

It is already clear from Fig 1 that HMXBs are unlikely to contribute
significantly to the total power output, kinetic or radiative, of
X-ray binaries (with the notable exceptions of SS 433 and Cyg X-3
which are powerful jet sources; see section 4). Simple calculations
bear this out and from this point on we do not consider HMXBs further.

\begin{figure}
\centerline{\epsfig{file=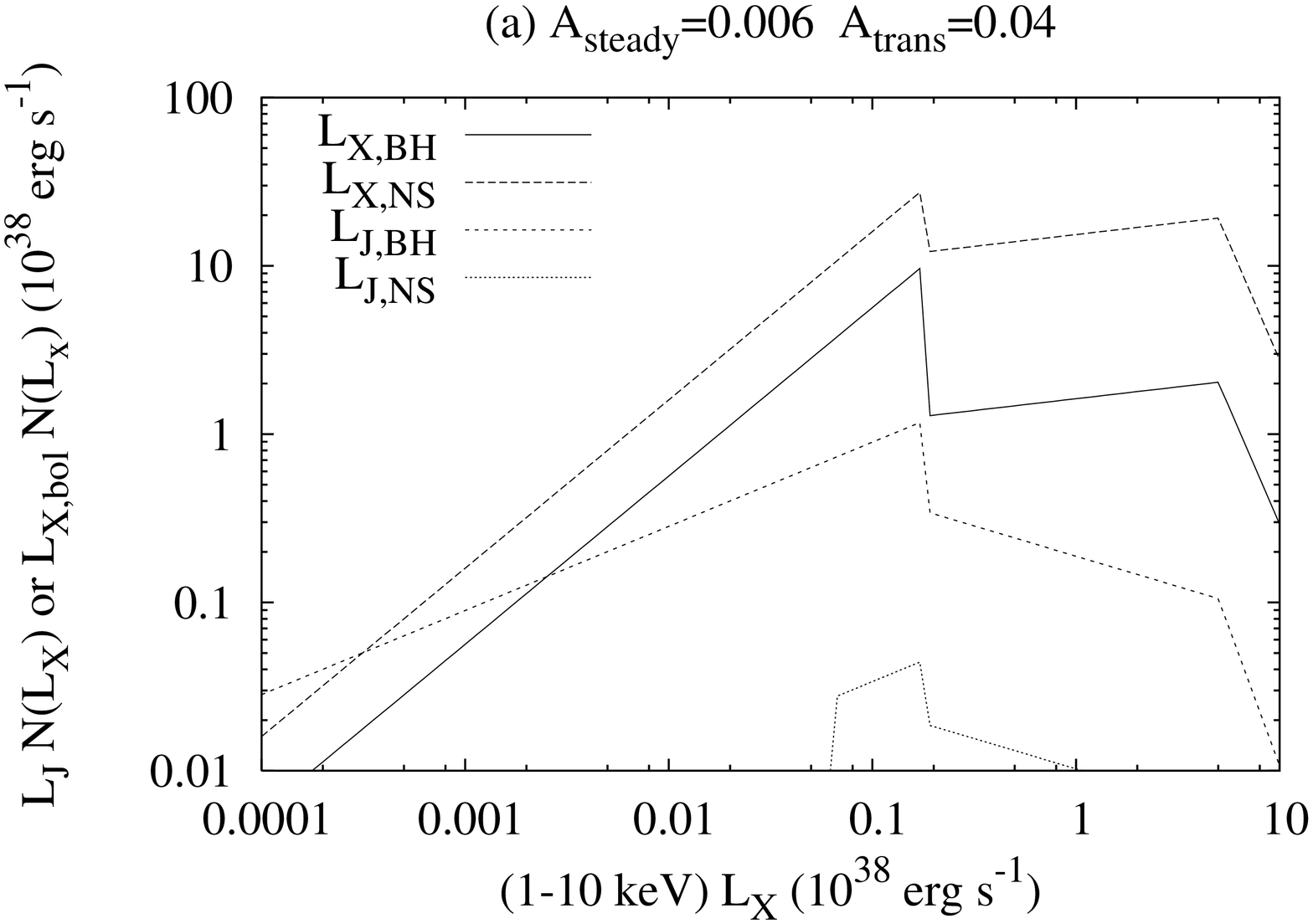, width=6cm, angle=270}}
\centerline{\epsfig{file=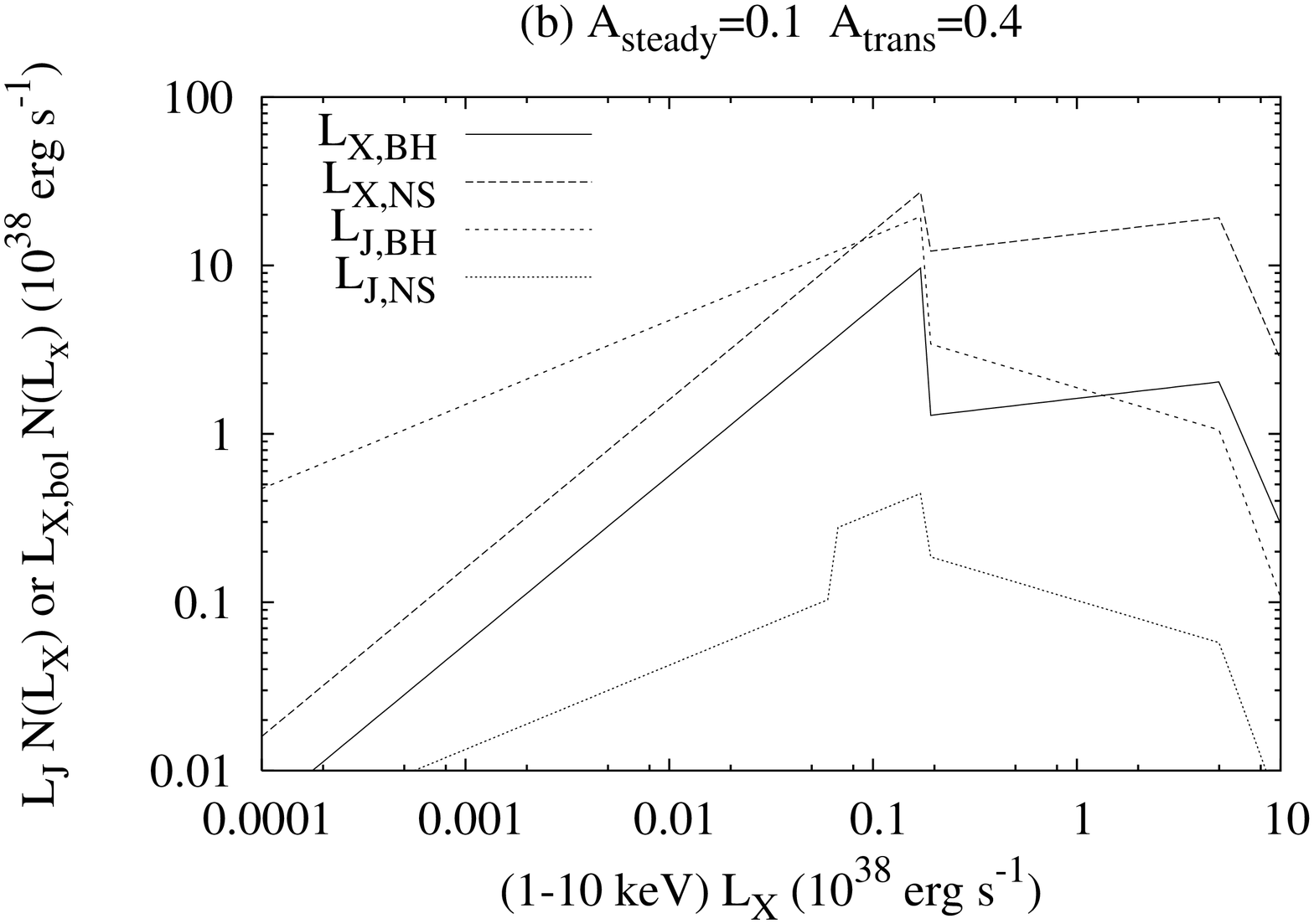, width=6cm, angle=270}}
\centerline{\epsfig{file=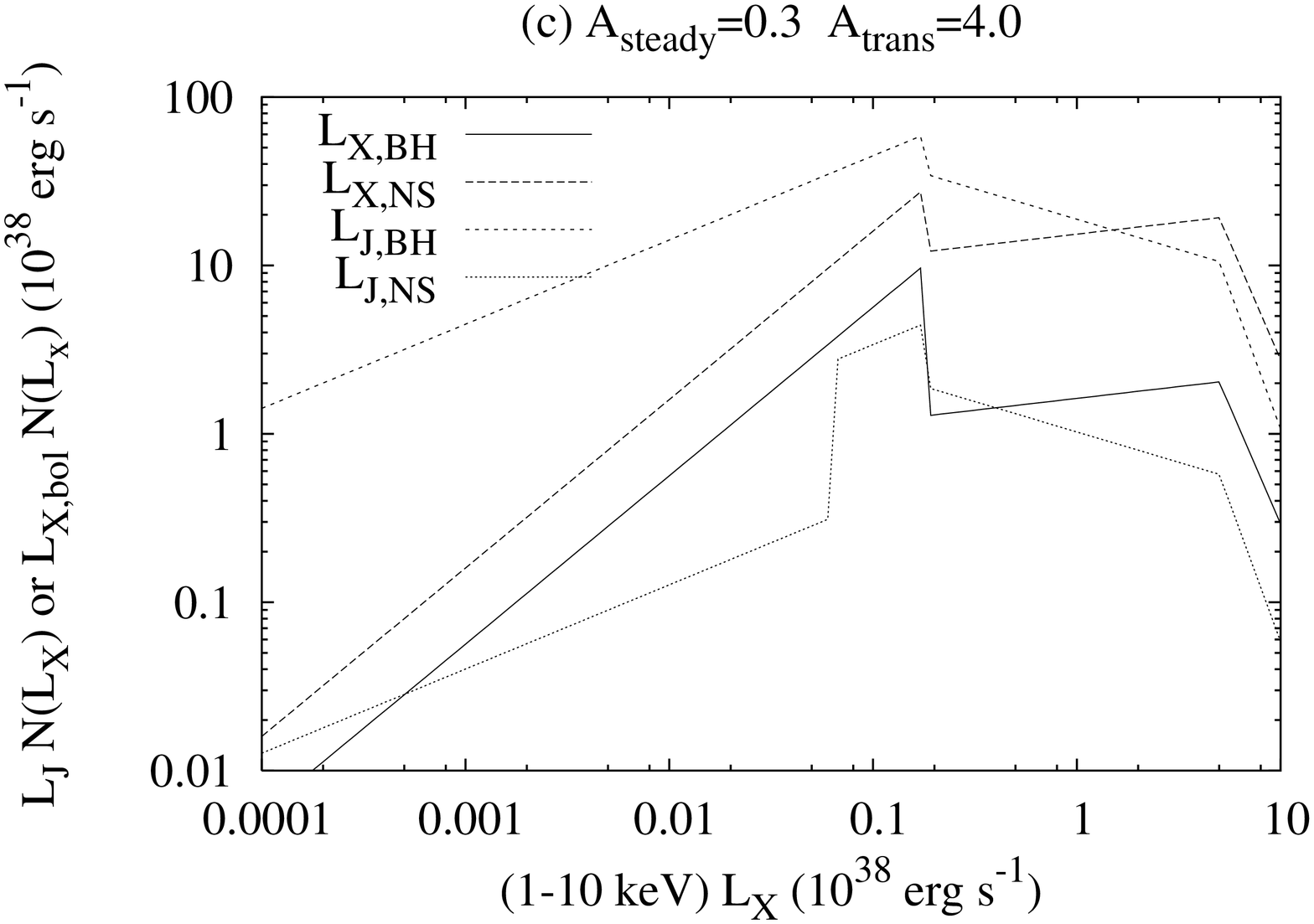, width=6cm, angle=270}}
\caption{The summed power output of all the Milky Way X-ray binaries, in radiation (X-rays) and kinetic power (jets), 
for three different sets of jet power normalisations, as a function of soft X-ray luminosity. The neutron stars
dominate the radiative output but the kinetic power output is dominated by the black holes.}
\end{figure}

\subsection*{Radiative and kinetic output of LMXBs}

In Fig 2 we present the integrated X-ray and jet (kinetic)
luminosities of LMXBs as a function of source X-ray luminosity, scaled
for the Milky Way, for three sets of values of jet power
normalisations $A_{\rm steady}$ and $A_{\rm trans}$. The shape of the
functions result from the multiplication of the LMXB XLF (equation
(1)) and the switch from steady to transient jets, which we fix to
occur at $1.9 \times 10^{37}$ for the black holes and $6 \times
10^{36}$ for the neutron stars, e.g. around 20\% of the Eddington
limit for each class of object. Some features are common
between the figures, and indeed independent of the selected values of
the jet normalisation, namely:

\begin{itemize}
\item{The radiative (X-ray) output is dominated by the more numerous NS LMXBs}
\item{The kinetic (jet) output is dominated by the BHC LMXBs, which produce more jet power at a given $L_{\rm X}$}
\end{itemize}

The integrated bolometric radiative (X-ray) output of the two classes of
objects is $L_{X, bol, NS} = 6.4 \times 10^{39}$ erg s$^{-1}$ and
$L_{X, bol, BHC} = 1.7 \times 10^{39}$ erg s$^{-1}$.

Moving on to individual figures, Fig 2(a) ('case A') considers the 'minimum' or
conservative estimates of the jet normalisation. In this scenario both
classes of object, NS and BHC, produce more integrated radiative than
kinetic ouput.  Even with these low values for the normalisation, a
transition from 'X-ray dominated' to 'Jet dominated' is apparent for
the BHCs (see Fender, Gallo \& Jonker 2003; Malzac, Merloni \& Fabian
2004 for a fuller discussions). The integrated BHC jet luminosity is 
$L_{\rm J} = 1.5 \times 10^{38}$ erg s$^{-1}$ (22\% from transient jets).

Fig 2(b) ('case B') considers the situation where the steady jets reach power
equipartition with the X-rays at the state transition, and dominate at
lower X-ray luminosities (see e.g. Livio, Pringle \& King 2003 for
support for this point of view). The transient jets have a
normalisation corresponding to the best fit from Fender, Belloni \&
Gallo (2004). In this scenario the peak of the kinetic power output
from the BHC LMXBs is comparable to the peak of the radiative power output
from the NS LMXBs. The integrated jet luminosity is
$L_{\rm J} = 2.0 \times 10^{39}$ erg s$^{-1}$ (19\% from transient jets).

Fig 2(c) ('case C') is perhaps the most 'extreme' case, in which the kinetic
output from the BHC LMXBs dominates the radiative output at
essentially all $L_{\rm X}$. The integrated jet luminosity is
$L_{\rm J} = 1.0 \times 10^{40}$ erg s$^{-1}$ (26\% from transient jets).

\section{Jets vs. Supernovae}

A 'standard' type-I or type-II supernova explosion is believed to
deposit around $10^{51}$ erg of kinetic energy into the surrounding
interstellar medium (e.g. Chevalier 1977; Korpi et al. 1999). For an
estimated rate of one such event per 100 years in the Milky Way, the
time-averaged kinetic power is $L_{SNe} = 3 \times 10^{41}$ erg
s$^{-1}$. Furthermore, Heinz \& Sunyaev (2002) estimate the cosmic ray
luminosity of the Milky Way as being approximately 10\% of this,
$L_{CR} = 4 \times 10^{40}$ erg s$^{-1}$.

Comparing these numbers with the three cases of jet power normalisation
considered above, we can see that 

\[
10^{-3} \la L_{\rm J} / L_{SNe} \la 0.03
\]

\[
10^{-2} \la L_{\rm J} / L_{CR} \la 0.3
\]

So it is clear that there may be enough power in the jets from X-ray
binaries to supply a sizeable fraction of the cosmic ray luminosity of
the Milky Way. There are many caveats to this, of course, but this
number is of considerable interest. If we consider 'case B' outlined
above to be the most likely, we find that the summed kinetic
luminosity of X-ray binary jets is $\ga$ 5\% of $L_{CR}$.

\subsection{Early-type galaxies}

So far we have only considered the situation for the Milky Way.
However, different types of galaxies have different ratios of X-ray
binaries to supernovae, which will affect the ratios calculated
above. Estimated supernova rates for different types of galaxies
are given in table 2 (from Cappellaro, Evans \& Tuatto 1999).

\begin{table}
\caption{Supernova rates in SNu for different galaxy types}
\begin{tabular}{cccc}
Galaxy type & Ia & Ib/c and II & Total\\
E-S0 & $0.18 \pm 0.06$ & $<0.03$  & $0.18 \pm 0.06$ \\
S0a-Sb & $0.18 \pm 0.07$ & $0.53 \pm 0.19$ & $0.72 \pm 0.21$ \\
Sbc-Sd & $0.21 \pm 0.08$ & $1.00 \pm 0.35$ & $1.21 \pm 0.37$ \\
\end{tabular}
\end{table}

Note that the total SNe rate in elliptical galaxies is about a
factor of six less than in late-type spirals such as the Milky Way
(due to the complete absence of SNe types Ib/c and II -- see e.g.
Cappellaro et al. 1999; van den Bergh, Li \& Fillipenko 2002). 

What do we expect of the X-ray binary, jet-producing, population ?
Gilfanov (2004) has shown that the population of LMXBs scales linearly
with the stellar mass of the galaxy, whereas for early-type galaxies
the HMXB population is expected to be essentially zero. We should note
that at least two of the most powerful jet sources, Cyg X-3 and SS
433, appear to be HMXBs (but in fact were not included in the
preceding analysis as all HMXB contributions were
disregarded). Nevertheless, GRS 1915+105 is a LMXB as are the seven
most luminous neutron star XRB jet sources (the Z sources plus Cir
X-1), as are all the BHC transients.  Based on this we may crudely
estimate that the jet power from 'outbursting' systems, per unit mass,
is reduced by about a factor of no more than two in ellipticals
compared to late-type spirals.

As an example we may take the case of NGC 4472, a large early-type
galaxy.  Gilfanov (2004) gives its stellar mass (the component which
is correlated with the number of LMXBs) as being 3.5 times that of the
Milky Way.  Considering the scaling of LMXB numbers with stellar mass,
as well as the possible 50\% reduction in the brightest sources due to
the lack of 'odd' HMXBs such as SS 433 and Cyg X-3, we would expect
the summed jet power in NGC 4472 to be about twice that in the Milky
Way. The total number of supernovae should also scale by mass, but is
reduced by a factor $\sim 6$ due to the lack of any type Ib/c or type
II events, and so would be around half of that in the Milky Way.
Therefore the ratio of $L_{\rm J} / L_{SNe}$ would be $\sim 4$ times greater
in this system, which may be considered to by typical for early-type
ellipticals. In this situation, 'case C' outlined above would correspond
to a total jet power input which was $\ga$ 15\% of the power output
of SNe. Even in the more conservative 'case B' the X-ray binary
jets would be injecting a significant amount of power into the ISM.

\section*{Conclusions and Caveats}

In this paper we have demonstrated that the integrated kinetic power
from X-ray binary jets in our galaxy is sufficient for them to be a
significant source of cosmic rays. Observations have already clearly
established that jets from powerful black hole transient XRBs can
accelerate electrons to very high (TeV) energies (e.g. Corbel et
al. 2002), but there remains little direct evidence for the
acceleration of protons or atomic nuclei to relativistic energies
(although heating of atomic nuclei to $\sim 10^7$K at multiple sites
in the jets of SS 433 has been demonstrated e.g. Marshall, Canizares
\& Schulz 2001 and references therein; Migliari, Fender \& Mendez
2002). However, unequivocal evidence for acceleration of the
non-leptonic component of cosmic rays in supernovae remnant shocks was
also very thin on the ground until recently (Aharonian et al. 2004), and
historically the argument was based upon energetics alone.

Given that the underlying physics of particle acceleration in SNR
shocks and XRB jets / jet-ISM interactions is likely to be similar, if
one process can accelerate protons to very high energies then presumably so
can the other. It should furthermore be noted that XRB jets almost
certainly spend a considerably longer phase moving with bulk
relativistic velocities than do SNR, and the variation of efficiency
of shock acceleration with bulk Lorentz factor is not well known.  So
it returns, at this stage, to simply a question of energy budgets.
Within the Milky Way, we conclude that the integrated jet power is
likely to be around 1\% of the rate of injection of energy from
supernovae, which is around 5--10\% of the cosmic ray luminosity,
comparable to the numbers estimated by Heinz \& Sunyaev (2002). There
exists a distinct possibility that these numbers are serious
underestimates. In addition, the total number of black holes in 
our galaxy accreting at very low levels remains highly uncertain, and
the contribution from such sources is difficult to quantify; this is
an area which needs some consideration.

The situation in the early-type galaxies, as illustrated with the
example of NGC 4472, is therefore even more striking. In this system,
with a mass 3.5 times that of the Milky Way, the total supernovae rate
is likely to be only half that of our galaxy, yet with a a population
of LMXBs which scales as the mass. Even if the energisation of cosmic
rays in our own galaxy is dominated by SNe, in such systems a
signficant fraction of the cosmic ray energy input should come from
X-ray binary jets.

Furthermore, by assuming, as suggested by Gilfanov (2004), that the
normalisation of the LMXB X-ray luminosity function is proportional to
a galaxy's stellar mass, we have most likely underestimated the
importance of LMXBs in giant elliptical galaxies and especially in the
central dominant (cD) galaxies of clusters; some scatter is found in
the relation of Gilfanov (2004), and in fact, the residuals are
correlated with the globular cluster specific frequencies of the
galaxies (White, Sarazin \& Kulkarni 2002).  Furthermore, it is known
that globular cluster specific frequencies are generally higher in the
most massive galaxies (e.g. McLaughlin, Harris \& Hanes 1993), and the
metal rich globular clusters, which are more likely to have X-ray
binaries (Kundu, Maccarone \& Zepf 2002) are even more prevelant in
these systems (e.g. Kundu \& Whitmore 2001) -- the fraction of the
X-ray sources in globular clusters clearly increases as one moves from
S0 to giant elliptical to cD galaxies (Maccarone, Kundu \& Zepf 2003).
As a result, the most massive galaxies are likely to have somewhat
higher $L_{LMXB}/L_{NIR}$ ratios than the Milky Way, making the
estimates here conservative.

There are of course many caveats and unknowns associated with the
entirety of the above discussion, as there are with the assumption of
SNe-dominated energisation.  These include the energetic contribution
from gamma-ray bursts and 'hypernovae' (e.g. Iwamoto et al. 1998),
although the rates for such events are likely to be orders of
magnitude less than those of core collapse supernovae (Podsiadlowski
et al. 2004). Furthermore, there exists a distinct possibility that
there is a poorly-known population of radiatively-faint sources with
powerful jets such as LS 5039 (Paredes et al. 2000).

To conclude, we argue that the input of kinetic energy into the ISM in
our galaxy may have a significant contribution from the jets of X-ray
binaries, although is still likely to be dominated by supernovae. In
early-type galaxies, which have no core collapse supernovae but a
comparable number of low-mass X-ray binaries per unit stellar mass,
the relative contribution will be much larger and should be considered
as potentially rivalling supernovae.  Given the high Lorentz factors
of the most powerful X-ray binary jets, their effect may be most
evident in its contribition to the production of cosmic rays. Finally,
the energetic contribution of these sources to the ISM will be in the
form of a much more distributed source than the rare, more energetic,
supernovae, which may have implications for the distribution of cosmic
rays throughout the galaxy, and be an important input for models of
their propagation through the galaxy (e.g. Strong, Moskalenko \&
Reimer 2004).

\section*{Acknowledgements}

RPF would like to thank Marat Gilfanov, Hans-Jakob Grimm, Christian
Kaiser, John Kirk and Ralph Wijers for useful conversations. 
Equally, TJM would like to thank Tiziana di Salvo, Chris Belczynski 
and Vicki Kalogera. RPF and TJM  would both like to thank Jeoren Homan for a
very useful contribution to our understanding of the typical duty
cycles of different states in X-ray outbursts.


\begin{thebibliography}{}


\bibitem[]{}
Aharonian F.A. et al., 2004, Nature, 432, 75

\bibitem[]{}
Blandford R.D., K\"onigl A., 1979, ApJ, 232, 34

\bibitem[]{}
Cappellaro E., Evans R., Turatto M., 1999, A\&A 351, 459

\bibitem[]{}
Chevalier R., 1977, ARA\&A, 15, 175

\bibitem[]{}
Corbel S., Fender R.P., Tzioumis A.K., Nowak M., McIntyre V., Durouchoux P., Sood R., 2000, A\&A, 359, 251

\bibitem[]{}
Corbel S. et al., 2001, ApJ, 554, 43

\bibitem[]{}
Corbel S., Nowak M., Fender R.P., Tzioumis A.K., Markoff S., 2003, A\&A, 400, 1007

\bibitem[]{}
Corbel S., Fender R.P., Tomsick J.A, Tzioumis A.K., Tingay S., 2004,
ApJ, 617, 1272

\bibitem[]{}
Elmegreen B.G., Scalo J., 2004, ARA\&A, 42, 211

\bibitem[]{}
Falcke H., Biermann P.L., 1996, A\&A, 308, 321

\bibitem[]{}
Fender R.P., 2001, MNRAS, 322, 31

\bibitem[]{} Fender R.P., 2005, in 'Compact Stellar X-Ray Sources',
eds. W.H.G. Lewin and M. van der Klis, Cambridge University Press,
{\bf (astro-ph/0303339)}

\bibitem[]{}
Fender R.P., Kuulkers E., 2001, MNRAS, 324, 923

\bibitem[]{}
Fender R.P., Gallo E., Jonker P.G., 2003, MNRAS, 343, L99

\bibitem[]{}
Fender R.P., Belloni T., Gallo E., 2004, MNRAS, 355, 1105

\bibitem[]{}
Fender R.P. et al., 1999, ApJ, 519, L165

\bibitem[]{}
Gallo E., Fender R.P., Pooley G.G., 2003, MNRAS, 344, 60
\bibitem[]{}
Gilfanov M., 2004, MNRAS, 349, 146

\bibitem[]{}
Grimm H.-J., Gilfanov M., Sunyaev R., 2002, A\&A, 391, 923

\bibitem[]{}
Grimm H.-J., Gilfanov M., Sunyaev R., 2003, MNRAS, 339, 793

\bibitem[]{}
Heinz S., Sunyaev R., 2002, A\&A, 390, 751

\bibitem[]{}
Heinz S., Sunyaev R., 2003, MNRAS, 343, L59

\bibitem[]{} Homan J., Belloni T., to appear in From X-ray Binaries to
  Quasars: Black Hole Accretion on All Mass Scales, ed. T. J.
  Maccarone, R. P. Fender, L. C. Ho (Dordrecht: Kluwer).
  (astro-ph/0412597)

\bibitem[]{}
Iwamoto K. et al., 1998, Nature, 395, 672

\bibitem[]{}
Korpi M.J., Brandenburg A., Shukurov A., Tuominen I., Nordlund A., 1999, ApJ, 514, L99

\bibitem[]{}
Kundu, A. \& Whitmore, B.C., 2001, AJ, 121, 2950 

\bibitem[]{}
Kundu, A., Maccarone, T.J. \& Zepf, S.E., 2002, ApJL, 574, 5

\bibitem[]{}
Liu, Q. Z., van Paradijs J., van den Heuvel E. P. J., 2000, A\&AS,
147, 25

\bibitem[]{}
Liu, Q. Z., van Paradijs J., van den Heuvel E. P. J., 2001, A\&A,
368, 1021

\bibitem[]{} McClintock J.E., Remillard R.A., 2005, in 'Compact
Stellar X-Ray Sources', eds. W.H.G. Lewin and M. van der Klis,
Cambridge University Press, {\bf (astro-ph/0306213)}

\bibitem[]{}
McLaughin, D.E., Harris, W.E. \& Hanes, D.A., 1993, ApJL, 409, 45

\bibitem[]{}
Maccarone, T.J., Kundu, A., \& Zepf, S.E., 2003, ApJ, 586, 814

\bibitem[]{}
McCray R., Snow T.P., 1979, ARA\&A, 17, 213

\bibitem[]{}
Malzac J., Merloni A., Fabian A.C., 2004, MNRAS, 351, 253

\bibitem[]{}
Markoff S., Falcke H., Fender R.P., 2001, A\&A, 372, L25

\bibitem[]{}
Marshall H.L., Canizares C.R., Schulz N.S., 2002, ApJ, 564, 941

\bibitem[]{}
Migliari M., Fender R., Mendez M., 2002, Science, 297, 1673

\bibitem[]{}
Migliari S., Fender R.P., Rupen M., Jonker P.G., Klein-Wolt M., Hjellming R.M., van der Klis M., 2003, 342, L67

\bibitem[]{}
Mirabel, I.F., Rodr\'\i guez, L.F, 1999, ARA\&A, 37, 409

\bibitem[]{}
Muno M.P., Belloni T., Dhawan V., Mogran E.H., Remillard R.A., Rupen M.P., ApJ, submitted (astro-ph/0411313)

\bibitem[]{}
Paredes J.M., Marti J., Ribo M., Massi M., 2000, Science, 288, 234

\bibitem[]{}
Podisalowski P., Mazzali P.A., Nomoto K., Lazzati D., Cappellaro E., 2004, ApJ, 607, L17


\bibitem[]{}
Portegies Zwart, S.F., Dewi, J. \& Maccarone, T., 2004, MNRAS, 355, 413

\bibitem[]{}
Strong A.W., Moskalenko I.V., Reimer O., 2004, ApJ, 613, 962

\bibitem[]{}
van den Bergh S., Li. W., Fillipenko A.V., 2002, PASP, 114, 820

\bibitem[]{}
White, R.E. III, Sarazin C.L. \& Kulkarni, S.R., 2002, ApJL, 571, 23


\end{thebibliography}
\end{document}